\documentclass[twocolumn,prb,10pt,showpacs,aps,footinbib]{revtex4}
\usepackage{graphicx}
\usepackage{color}
\usepackage{multirow}
\usepackage{parskip}

\begin{document}

\title{GW quasiparticle band structures of stibnite, antimonselite, \\ 
bismuthinite, and guanajuatite} 

\author{Marina R. Filip}
\author{Christopher E. Patrick}
\author{Feliciano Giustino}
\affiliation{Department of Materials, University of Oxford, Parks Road,
Oxford OX1 3PH, United Kingdom}
\begin{abstract}
We present first-principles calculations of the quasiparticle band structures of
four isostructural semiconducting metal chalcogenides A$_2$B$_3$ (with A = Sb,
Bi and B = S, Se) of the stibnite family within the G$_0$W$_0$ approach. We perform 
extensive convergence tests and identify a sensitivity of the quasiparticle 
corrections to the structural parameters and to the semicore $d$ electrons. Our calculations
indicate that all four chalcogenides exhibit direct band gaps, if we 
exclude some indirect transitions marginally below the direct gap. 
Relativistic spin-orbit effects are evaluated for the Kohn-Sham band structures,
and included as scissor corrections in the quasiparticle band gaps. 
Our calculated band gaps are 1.5~eV (Sb$_2$S$_3$), 1.3~eV 
(Sb$_2$Se$_3$), 1.4~eV (Bi$_2$S$_3$) and 0.9~eV (Bi$_2$Se$_3$). 
By comparing our calculated gaps with the ideal Shockley-Queisser value 
we find that all four chalcogenides are promising as light sensitizers for nanostructured photovoltaics.
\end{abstract}

\maketitle

\section{Introduction}

The development of sustainable energy solutions based on scalable processes and
non-toxic materials constitutes a key priority in the current scientific research agenda, and in this
area nanostructured energy-harvesting solar and thermoelectric devices are playing a lead role. 
Recently there has been a surge of interest in devices using semiconducting metal 
chalcogenides of the stibnite family. For example recent studies have demonstrated
the potential of these semiconductors both in photovoltaics 
applications,\cite{Chang2010, Chang2012, Konstantatos2012, Guijarro2012} and in
thermoelectric generators.\cite{Mehta2010} 

In the area of nanostructured photovoltaics semiconducting metal chalcogenides
have successfully been used to replace the inorganic dye in dye-sensitized solar
cells,\cite{O'Regan1991} leading to the development of solid-state semiconductor-sensitized
solar cells.\cite{Hodes2008,Chang2010}
In these devices thin layers or nanoparticles of the semiconducting chalcogenides
act as light absorbers, and upon photoexcitation they transfer an electron to the acceptor
(typically TiO$_2$) and a hole to the hole-transporter (for example a conducting polymer).
The record efficiency within this class of devices is 5.1\% and was obtained using
stibnite (Sb$_2$S$_3$) as semiconductor sensitizer.\cite{Chang2010}

A recent atomistic computational study of photovoltaic interfaces for semicondictor-sensitized
solar cells pointed out that, in addition to stibnite, the other members of the 
stibnite mineral family, namely antimonselite (Sb$_2$Se$_3$), bismuthinite (Bi$_2$S$_3$), 
and guanajuatite (Bi$_2$Se$_3$), exhibit optical properties similar to stibnite
and should be considered as potential candidates for novel semiconductor sensitizers.\cite{Patrick2011}
Using density-functional calculations and empirical scissor corrections of the band gaps, 
in Ref.~\onlinecite{Patrick2011} it was found that stibnite and antimonselite
should form type-II heterojunctions with TiO$_2$, while bismuthinite should form
a type-I heterojunction and hence would not be able to transfer electrons to TiO$_2$.
These theoretical predictions have recently been confirmed by the experimental
investigations of Refs.~\onlinecite{Guijarro2012, Lutz2012}, thereby providing
a motivation for further studies and for the more sophisticated analysis presented 
in this work.

\begin{figure}[b]
  \begin{center}
  \includegraphics[width=0.8\columnwidth]{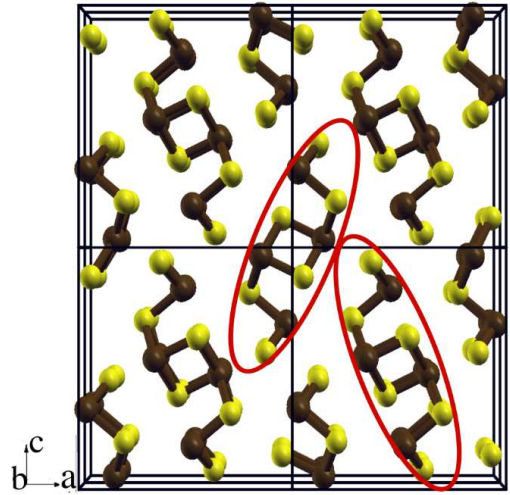}
  \caption{
  Ball-and-stick model of A$_2$B$_3$ semiconducting metal chalcogenides of the stibnite family,
  with A standing for Sb or Bi (brown), and  B for S or Se (yellow). The two inequivalent 
  (A$_4$B$_6$)$_n$ ribbons in the unit cell are highlighted in red, and the perspective view 
  is along the direction of the ribbons.
 }
  \label{fig:cryst}
  \end {center}
  \end{figure}

The four minerals of the stibnite family crystallize in an orthorhombic structure
consisting of parallel one-dimensional (A$_4$B$_6$)$_n$ ribbons,
with A = Sb, Bi and B = S, Se. A ball-and-stick model of this structure is shown 
in Fig.~\ref{fig:cryst}. 
Besides its natural occurrence in mineral form, stibnite can be synthesised using 
a variety of low-cost fabrication techniques.\cite{Bhosale1994, Versavel2007, Lokhande2002, 
Maghraoui-Meherzi2010, Cademartiri2008, Malakooti2008, Ruhle2010, 
Han2011, Cademartiri2012} Using these techniques it is possible to obtain a good degree
of crystallinity,\cite{Rincon1998,Perales2007} to control dimensionality,\cite{Ruhle2010, Bao2007, Mehta2010}
and to tune the optical properties.\cite{Vedeshwar1995, Karguppikar1987, Ruhle2010, Deng2009}
Semiconductors of the stibnite family have also been synthesized in various nanostructured forms.
For example Refs.~\onlinecite{Cademartiri2008, Malakooti2008, Mehta2010} and 
Refs.~\onlinecite{Mehta2010,Deng2009} reported nanowires and nanotubes, respectively, 
of stibnite, antimonselite and bismuthinite.
Nanowires of stibnite were found to exhibit enhanced ferroelectric and piezoelectric properties 
as compared to their bulk counterpart.\cite{Varghese2012}
Nanowires and nanotubes of antimonselite were found to exhibit conductivities much higher
than their bulk counterpart,\cite{Mehta2010} and are being considered for thermoelectric
applications. In the case of bismuthinite, Ref.~\onlinecite{Cademartiri2008} reported nanowires 
with diameters as small as 1.6~nm, corresponding to a transverse size of only two ribbons. 
The rhombohedral phase of Bi$_2$Se$_3$ has been investigated extensively
since this compound is a prototypical topological insulator.\cite{Zhang2009} However
to the best of our knowledge little is known about orthorhombic Bi$_2$Se$_3$, i.e.\ guanajuatite,
which is stable only at high temperature and pressure.\cite{Abateva1973, Okamoto1994}

The band gaps of stibnite, antimonselite and bismuthinite have been measured extensively 
via optical absorption experiments. The band gap of stibnite ranges between 
1.42-1.78~eV.\cite{Mahanty1997,Zawawi1998}
For antimonselite Ref.~\onlinecite{Torane1999} reported a direct gap of 1.55~eV,
while Ref.~\onlinecite{Rodriguez-Lozcano2005} gave an indirect gap between 1-1.2~eV.
The measured band gap of bismuthinite is 1.38-1.58~eV.\cite{Pradeep1991,Mahmoud1997,Yesugade1995}
The spread in the measured gaps can be attributed to the
different preparation conditions used, yielding different degrees of polycrystallinity
and even amorphous samples in some cases, and also different stoichiometries.
In addition all these compounds exhibit closely lying direct
and indirect transitions (cf.\ Fig.~\ref{fig:bandstruct} below), thereby complicating
the assignment of the nature of the optical gap.

All four minerals of the stibnite family have been investigated in detail
using density-functional theory (DFT) calculations. The electronic properties of
these compounds have been studied in Refs.~\onlinecite{Caracas2005, Vadapoo2011,Larson2002, 
Koc2012, Sharma2010, Sharma2012, Nasr2011, Patrick2011},
and the elastic and optical properties have been calculated in Ref.~\onlinecite{Koc2012}.

A comparison of the theoretical studies published so far shows some inconsistencies 
in the calculated band gaps, for example the values reported for stibnite are in
the range 1.18-1.55~eV.\cite{Caracas2005, Patrick2011, Vadapoo2011rib, Koc2012}
As expected all the calculated DFT gaps underestimate the measured band gaps.
To the best of our knowledge only one work\cite{Vadapoo2011rib} reported a calculation 
of the quasiparticle band gap of stibnite and antimonselite within the 
GW approximation.\cite{Hedin1965} The electronic structure of the rhombohedral Bi$_2$Se$_3$ has 
also been explored within the GW approach \cite{Yazyev2012}.

Within this context there exists a need for detailed and reproducible calculations
of the electronic structure of stibnite and related compounds based on state-of-the-art 
quasiparticle techniques. In line with this need the goal of the present work is to report 
a systematic and reproducible study of the quasiparticle band structures of all four A$_2$B$_3$ 
semiconducting metal chalcogenides of the stibnite family. An emphasis is placed on convergence 
tests and on the sensitivity of the quasiparticle corrections to the structural parameters,
the inclusion of semicore $d$ states in the calculations, and relativistic
effects.

Our calculated band gaps are 1.5 eV (Sb$_2$S$_3$),
1.3 eV (Sb$_2$Se$_3$), 1.4 eV (Bi$_2$S$_3$) and 0.9 eV (Bi$_2$Se$_3$).
By inspection of the band structures
we infer that all four compounds have direct band gaps, although in most cases an indirect
transition just below the direct gap (within 0.1~eV) is also possible.
The inclusion of semicore electrons in the calculations is found to modify 
the band gaps by 0.1-0.2~eV. In addition we find that the gaps are rather sensitive
to the lattice parameters, as they change by up to 0.3~eV when the lattice parameters
are taken from experiment or fully optimized within DFT. Relativistic corrections 
are found to be essentially negligible for Sb$_2$S$_3$ and Sb$_2$Se$_3$, while in the case
of Bi$_2$S$_3$ and Bi$_2$Se$_3$ the band gaps decrease by 0.3-0.4~eV upon inclusion
of spin-orbit coupling.

The manuscript is organized as follows. In Sec.~\ref{sec:Methodology} we describe the computational
methodology and the convergence tests for the GW calculations. In Sec.~\ref{sec:Results} we present
our main results, including quasiparticle band structures and band gaps. In Sec.~\ref{sec:Discussion}
we discuss our findings in relation to the photovoltaics applications of the materials considered 
in this work. In Sec.~\ref{sec:Conclusions} we summarize our results and present our conclusions.

\section{Methodology}\label{sec:Methodology}

\subsection{DFT calculations}\label{sec.dft}

All DFT calculations are performed using the \texttt{Quantum ESPRESSO} package.\cite{QE-2009} 
The calculations are based on the local density approximation (LDA) to 
DFT.\cite{Perdew1981, Ceperley1980} 

Only valence electrons are explicitly 
described, and the core-valence interaction is taken into account by means of Troullier-Martins scalar relativistic pseudopotentials\cite{Martins1991} generated using the \texttt{fhi98} code.\cite{Fuchs1999}
In the cases of S (Se) the $3s^23p^4$ ($4s^24p^4$) electrons are included in the valence as usual.
For Sb and Bi we generate two sets of pseudopotentials, one set with five electrons in the
valence, i.e. $5s^25p^3$ and $6s^26p^3$ respectively, and one set with additional semicore 
$4d^{10}$ and $5d^{10}$ electrons, respectively.

The electronic wavefunctions are expanded in planewaves basis sets with kinetic energy cutoffs
of 70 Ry (Sb$_2$Se$_3$, Bi$_2$Se$_3$) and 90 Ry (Sb$_2$S$_3$, Bi$_2$S$_3$) for the
calculations without semicore states, and 100 Ry (Bi$_2$S$_3$, Bi$_2$Se$_3$) and 
130 Ry (Sb$_2$S$_3$, Sb$_2$Se$_3$) when semicore states are included.
In each case considered the selected cutoff yields a total energy converged to within 2 meV/atom.
All self-consistent calculations are carried out using a 8$\times$8$\times$8 
Brillouin zone mesh centered at $\Gamma$, corresponding to 170 irreducible points for Sb$_2$S$_3$ and Bi$_2$S$_3$, and 260 points for Sb$_2$Se$_3$ and Bi$_2$Se$_3$.

We perform full geometry optimizations of the lattice parameters and the atomic positions
in each case, both with or without semicore $d$ states.
All structural optimizations are performed using 4$\times$8$\times$4 $\Gamma$-centered Brillouin
zone meshes.

\subsection{Crystal structure}

Stibnite (Sb$_2$S$_3$), antimonselite (Sb$_2$Se$_3$), bismuthinite (Bi$_2$S$_3$),
and guanajuatite (Bi$_2$Se$_3$) all crystallize in the same orthorhombic lattice and belong
to the Pnma 62 space group.\cite{Caracas2005} Each unit cell contains 20 atoms,
whose coordinates can be generated by applying the symmetry operations of the
crystallographic group to a set of 5 atomic coordinates.
Figure~\ref{fig:cryst} shows a ball-and-stick representation of these A$_2$B$_3$ structures.
The structural parameters were measured by Refs.~\onlinecite{Bayliss1972, Voutsas1985, 
Kanishcheva1981, Abateva1973} and are reported in Ref.~\onlinecite{Caracas2005}.

As the crystal structure consists of a bundle of relatively well separated ribbons,
it is convenient to separate the cohesive energy into intra-ribbon and inter-ribbon
components. The intra-ribbon cohesive energy is calculated
as the difference between the total energy of one ribbon and the total energies of its
consituent atoms. The inter-ribbon cohesive energy is evaluated as the difference between the
total energy of the unit cell and twice the total energy of one ribbon in isolation
(each unit cell contains two ribbons).

\subsection{Quasiparticle calculations}

We calculate the quasiparticle energies within many-body perturbation theory using the GW 
method,\cite{Hedin1965, Hybertsen1986, Aulbur1999, Arya1998, Onida2002} 
as implemented in the \texttt{SaX} code.\cite{MartinSamos2009}
The GW self-energy is evaluated in the G$_0$W$_0$ approximation as $\Sigma = i G_0 W_0$.
Here $G_0$ denotes the electron Green's function defined by the Kohn-Sham 
eigenstates $\psi_{n\mathbf{k}}(\mathbf{r})$ and eigenvalues $\epsilon_{n\mathbf{k}}$ corresponding
to the band index $n$ and the wavevector $\mathbf{k}$, and $W_0$ represents the screened 
Coulomb interaction calculated in the random phase approximation.\cite{Hedin1969,Hybertsen1986}
The quasiparticle energies $E_{n\mathbf{k}}$ are obtained as:\cite{Hybertsen1986}
  \begin{equation}
  E_{n\mathbf{k}} = \epsilon_{n\mathbf{k}} + Z_{n\mathbf{k}} 
  \langle\psi_{n\mathbf{k}}|\Sigma(\epsilon_{n\mathbf{k}}) - V_{xc}|\psi_{n\mathbf{k}}\rangle,
  \label{eq:qp}
  \end {equation}
where $E_{n\mathbf{k}}$ is the quasiparticle energy, $Z_{n\mathbf{k}}$ the associated
quasiparticle renormalization, and $V_{xc}$ is the exchange and correlation potential. 

The self-energy can be written as the sum of a bare exchange contribution $\Sigma_x$ 
and a correlation contribution $\Sigma_c$: $\Sigma=\Sigma_x+\Sigma_c$.
The exchange part does not depend explicitly on the excitation energy
and reads:\cite{Giustino2010} 
  \begin{equation}
  \Sigma_x(\mathbf{r},\mathbf{r'}) = -\!\!\!\!\sum_{n\in {\rm occ},\mathbf{k}}\!
    \psi_{n\mathbf{k}}^*(\mathbf{r}) \psi_{n\mathbf{k}}^{\phantom{*}}(\mathbf{r'}) v(\mathbf{r},\mathbf{r'}),
  \label{eq:sx}
  \end{equation}
where the sum is over occupied states and $v$ represents the bare Coulomb interaction.
This contribution to the quasiparticle correction is sensitive to the overlap between
Kohn-Sham wavefunctions regardless of their energy. As a result the use of semicore states 
can have significant effect on the calculations, as shown in Refs.~\onlinecite{Rohlfing1995, Tiago2004, 
Umari2012}. This aspect will be discussed in detail in Sec.~\ref{sec.qp-corrections}.

The energy-dependence of the correlation contribution $\Sigma_c$ arising from the
dynamically-screened Coulomb interaction is described via the Godby-Needs plasmon-pole model.\cite{Godby1989}
We use a plasmon-pole energy of 1 Ry for all materials, similar to the energy of
the peaks in the calculated electron energy loss spectra.

\begin{figure*}
  \begin{center}
  \includegraphics[width=0.8\textwidth]{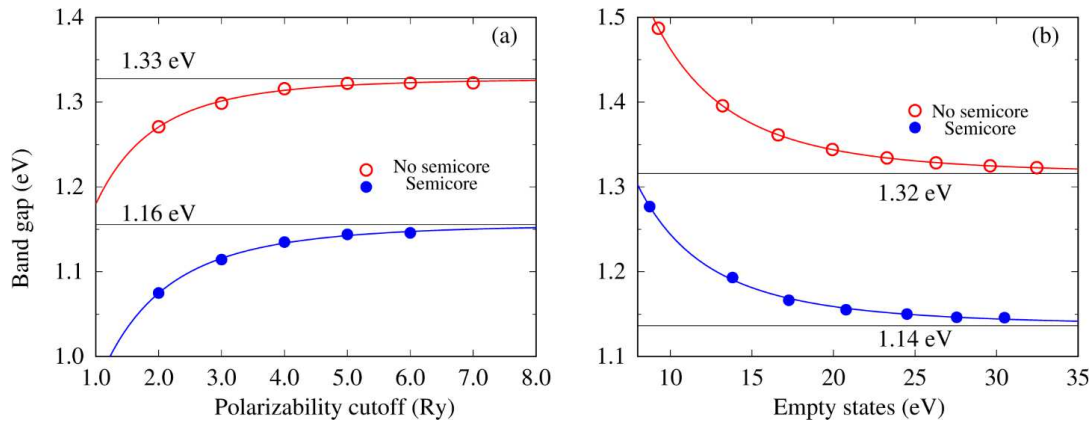}
  \caption{
  (color online)
  Convergence tests for the quasiparticle band gap of antimonselite. (a) Calculated $G_0W_0$
  band gap as a function of the polarizability cutoff, for calculations
  without (red circles) or with (blue disks) semicore states. The solid
  lines correspond to the fits obtained from Eq.~(\ref{eq:fit}). We find $a_0$=1.33~eV
  and 1.16~eV for calculations without and with semicore electrons, respectively. 
  (b) Same as in (a), with the band gap reported as a function of the largest energy
  denominator used for calculating the polarizability. In this case we find
  $a_0$=1.32/1.14~eV for calculations without/with semicore electrons.All calculations
  were performed using optimized lattice parameters.}
  \label{fig:conv}
  \end{center}
  \end{figure*}

Since the computational efforts for achieving convergence in $\Sigma_c$ and $\Sigma_x$
are very different owing to the necessity of evaluating unoccupied states for $\Sigma_c$,
we perform separate convergence tests for these two components.
For the exchange contribution we use kinetic energy cutoffs of 75~Ry and 100~Ry
for calculations without and with semicore electrons, respectively.
For the correlation contribution we perform convergence tests by calculating
the band gap at various kinetic energy cutoffs up to 7~Ry
for the polarizability.
Figure~\ref{fig:conv}(a) shows that the band gap is converged within 0.05~eV
already for a cutoff of 5~Ry.
The dependence of the band gap on the polarizability cutoff shows the same
trend for calculations with or without semicore states. This is consistent
with the expectation that the effect of semicore states in $\Sigma_c$ should be small.\cite{Rohlfing1995}
Based on the data of Fig.~\ref{fig:conv}(a), in the following
we use a polarizability cutoff of 7~Ry for calculations without semicore electrons, 
and of 6~Ry for the more demanding calculations including semicore states.
In Fig.~\ref{fig:conv}(b) we show the convergence of the band gap of antimonselite
with respect to the energy of the highest unoccupied state included in the
polarizability. 
Based on the trend in this figure we set the number of unoccupied states to 224 and 264 for
calculations with and without semicore, corresponding to a maximum energy denominator of 35~eV.
Both $G_0$ and $W_0$ are calculated on uniform and $\Gamma$-centered 2$\times$6$\times$2 
Brillouin-zone meshes.

In order to estimate the accuracy of our quasiparticle corrections with respect to 
the above convergence parameters we follow the approach of Ref.~\citenum{Sharifzadeh2012}. 
In this approach the dependence of the band gap on a given convergence parameter
is fitted by a simple function in order to extract a ``best-guess'' asymptotic limit.
This asymptotic limit is then taken to represent the converged gap.
In this work we tentatively approximate gap vs.\ cutoff curves using the following
function:
  \begin{equation}
  E^{\mathrm{QP}}_{\mathrm{gap}} = a_0 + a_1(x-a_2)^{-1/a_3},
  \label{eq:fit}
  \end{equation}
where $E_{\mathrm{gap}}^{\mathrm{QP}}$ is the quasiparticle band gap, $x$ is the convergence 
parameter (i.e.\ the polarizability cutoff or the largest energy denominator) and $a_0,\dots,a_3$ 
are fitting parameters. 
While Eq.~(\ref{eq:fit}) is largely arbitrary, this choice reflects the expectation that
the gap will converge faster than $1/x$ owing to the damping introduced by the matrix elements
in the Adler-Wiser polarizability.\cite{Adler1962, Wiser1963} 
Figure~\ref{fig:conv} shows that the fitting curves obtained for stibnite describe
rather accurately the calculated data points, therefore it is reasonable to assume that
the parameter $a_0$ obtained from the fit should provide a good estimate of the converged
gap. By repeating this procedure for all four compounds Sb$_2$S$_3$, Sb$_2$Se$_3$, Bi$_2$S$_3$,
and Bi$_2$Se$_3$ we find that the convergence parameters described above yield
band gaps which differ by less than 0.05~eV from the corresponding asymptotic values.

\subsection{Spin-orbit coupling}
Owing to the high atomic numbers of Bi and Sb, it is important to check the role
of spin-orbit coupling (SOC) in the electronic structure of
semiconductors of the stibnite family. In this work we evaluate SOC effects at the
DFT level, by constructing a set of fully-relativistic Troullier-Martins pseudopotentials
including semicore $d$ states.
The pseudopotentials are generated using the \texttt{ld1.x} program of the
\texttt{Quantum ESPRESSO} package. 
We checked that the planewaves cutoffs described in Sec.~\ref{sec.dft} 
for scalar-relativistic calculations are also appropriate for these fully-relativistic
pseudopotentials.
For S and Se relativistic 
effects are not expected to be significant, and scalar-relativistic pseudopotentials are used
throughout. 
We determine the spin-orbit corrections to the band gaps by taking the differences
between self-consistent calculations using the fully-relativistic pseudopotentials 
with or without noncollinear magnetism.\cite{Oda1998}
We then apply these differences as scissor corrections to the corresponding quasiparticle band gaps
obtained from scalar relativistic calculations.

\section{Results}\label{sec:Results}

\subsection{Structural parameters}

Table~\ref{tb:latconst} shows the comparison between our calculated lattice parameters
and experiment. As expected the use of the local density approximation leads to a general
underestimation of the experimental parameters. Interestingly, while
such underestimation does not exceed 1.1\% along the direction of the (A$_4$B$_6$)$_n$ ribbons
($b$ parameter in Table~\ref{tb:latconst}, cf.\ Fig.~\ref{fig:cryst}),
the deviation can reach up to 4.2\% in the direction perpendicular to the ribbons
($a$ and $c$ parameters in Table~\ref{tb:latconst}).
We tentatively assign this behavior to the fact that inter-ribbon forces are likely 
to include non-negligible van der Waals components, and hence are not described correctly within 
the LDA. 

Inspection of the calculated cohesive energies seems to support this possibility.
Indeed we obtain intra-ribbon cohesive energies of 3.9~eV/atom (Sb$_2$S$_3$), 3.5~eV/atom (Sb$_2$Se$_3$)
3.6~eV/atom (Bi$_2$S$_3$) and 3.3~eV/atom (Bi$_2$Se$_3$).
The inter-ribbon cohesive energy are one order of magnitude smaller,
0.2~eV/atom (Sb$_2$S$_3$ and Sb$_2$Se$_3$) and 0.3~eV/atom (Bi$_2$S$_3$ and Bi$_2$Se$_3$).

We also performed additional calculations
of the structural parameters using the van der Waals functional of Ref.~\onlinecite{Dion2005}.
The lattice parameters calculated using the vdW functional
overestimate the experimental values by up to 6.9\% along the directions perpendicular
to the ribbons, while along the ribbons the calculated parameters are in agreement with
experiment (within 0.3\%). Similar trends have been observed in calculations
on graphite and boron nitride in Ref.~\onlinecite{Langreth2005}. These results indicate that for semiconductors
of the stibnite family the use of a van der Waals functional does not improve the agreement
of the calculated structural parameters with experiment. 

In order to take into account the differences between calculated and experimental lattice
parameters, in the following we report quasiparticle calculations obtained using either the DFT/LDA
structure or the experimental structure.

  \begin{table*}
  \begin{center}
  \begin{tabular}{rrrrrrrrrrrr}
  \hline
  \hline
  \\[-8pt]
  & \multicolumn{3}{c}{Experiment}&& \multicolumn{3}{c} {Calc. w/o semicore} 
     &&\multicolumn{3}{c} {Calc. with semicore} \\
  \\[-8pt]
  \hline
  \\[-8pt]
  & $a$ & $b$ & $c$ && $a$ & $b$ & $c$ && $a$ & $b$ & $c$\\
  \\[-8pt]
  \multicolumn{1}{l}{Sb$_2$S$_3$} & 11.311$^a$ & 3.836$^a$ & 11.229$^a$ &
                                  & 11.036 & 3.795 & 10.753 && 11.087 & 3.838 & 10.834 \\
   & &					&				&& 			-2.4\% & 
   -1.1\% & -4.2\% && -2.0\% & 0.1\% & -3.5\% \\
  \\[-8pt]
  \multicolumn{1}{l}{Sb$_2$Se$_3$} & 11.794$^b$ & 3.986$^b$ & 11.648$^b$ &
                                  & 11.609 & 3.952 & 11.213 && 11.646 & 3.989 & 11.287 \\
   & &					&				&&			-1.6\% & 
   -0.9\% & -3.7\% && -1.3\% & 0.1\% & -3.1\% \\
  \\[-8pt]
  \multicolumn{1}{l}{Bi$_2$S$_3$} & 11.305$^c$ & 3.981$^c$ & 11.147$^c$ &
                                  & 11.227 & 3.999 & 11.001 && 11.030 & 3.949 & 10.853 \\
   & & 					& 				&&			-0.7\% & 
   0.5\% & -1.3\% && -2.4\% & -0.8\% & -2.6\%\\
  \\[-8pt]
  \multicolumn{1}{l}{Bi$_2$Se$_3$} & 11.830$^d$ & 4.090$^d$ & 11.620$^d$ &
                                        & 11.767 & 4.141 & 11.491 && 11.609 & 4.099 & 11.374 \\
   & & 	 				&				&&			-0.5\% & 
   1.3\% & -1.1\% && -1.9\% & 0.2\% & -2.1\% \\
  \\[-10pt]
  \hline
  \hline
  \footnotesize
  $^a$ Ref.~\citenum{Bayliss1972}.\\
  \footnotesize
  $^b$ Ref.~\citenum{Voutsas1985}.\\
  \footnotesize
  $^c$ Ref.~\citenum{Kanishcheva1981}.\\
  \footnotesize
  $^d$ Ref.~\citenum{Abateva1973}.
  \end{tabular}
  \caption{\label{tb:latconst} Comparison between the calculated DFT/LDA lattice parameters 
  of stibnite, antimonselite, bismuthinite, and guanajuatite and experiment (all values are
  given in \AA). The percentile deviation from experiment is indicated in each case.}
  \end {center}
  \end{table*}
  \begin{table}[h]
  \begin{center}
  \begin{tabular}{lccccc}
  \hline
  \hline
  \\[-8pt]
    & \multicolumn{2}{c}{Mimimum gap\phantom{aaa}} && \multicolumn{2}{c}{Direct gap\phantom{aaa}} \\
   & w semicore & w/o semicore && w semicore & w/o semicore \\
  \\[-8pt]
  \hline
  \\[-8pt]
  Sb$_2$S$_3$  & 1.19 & 1.21 && 1.26 & 1.27 \\
  Sb$_2$Se$_3$ & 0.84 & 0.86 && 0.84 & 0.86 \\
  Bi$_2$S$_3$  & 1.25 & 1.24 && 1.28 & 1.27 \\
  Bi$_2$Se$_3$ & 0.85 & 0.86 && 0.99 & 0.99 \\
\\[-10pt]
  \hline
  \hline
  \end{tabular}
  \caption{\label{tb:dftgap} 
  Comparison between the minimum band gaps and the direct band gaps
  of stibnite, antimonselite, bismuthinite, and guanajuatite, as obtained
  from DFT/LDA. In these calculations we use the experimental lattice parameters.
  All values are in units of eV.
  }
  \end {center}
  \end{table}

\subsection{DFT/LDA band structures}

\begin{figure*}
  \begin{center}
  \includegraphics[width=0.8\textwidth]{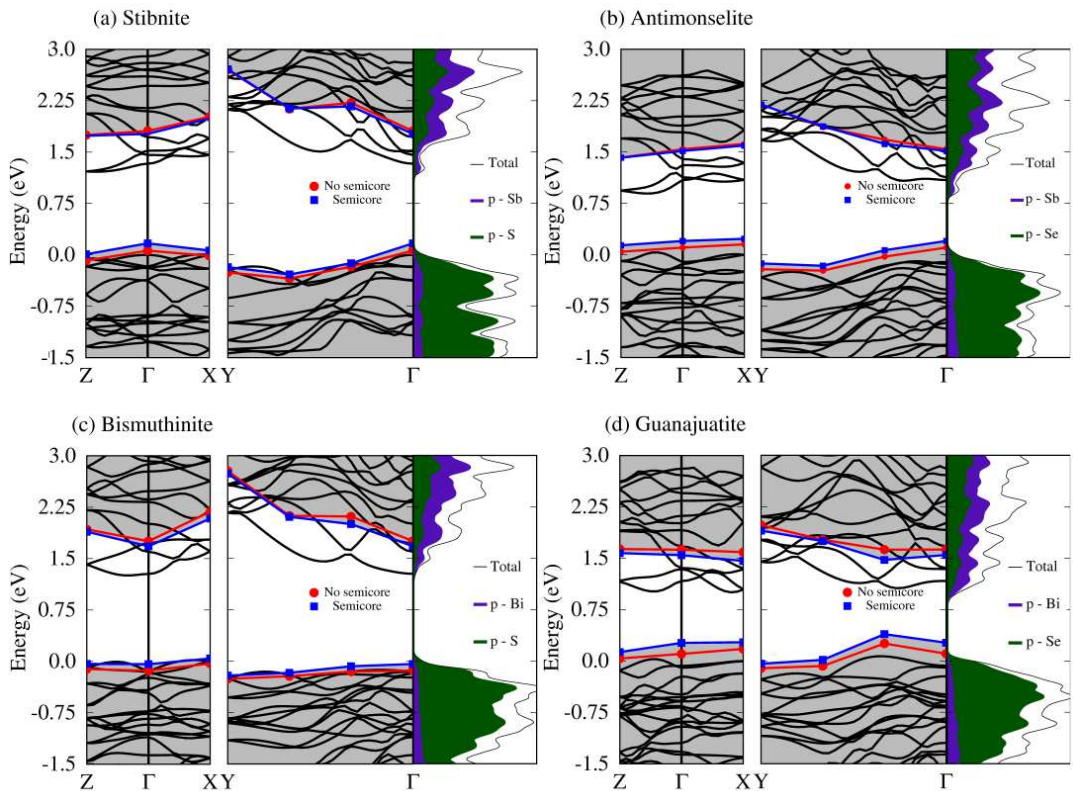}
  \caption{(color online)
  Band structures of (a) stibnite, (b) antimonselite, (c) bismuthinite, and (d) guanajuatite
  calculated using DFT/LDA, experimental lattice parameters, and without semicore electrons 
  (black solid lines), as well as corresponding density of states (DOS, black dashed lines). 
  The contributions to the DOS from the $p$ states of S and Se (Sb and Bi) are indicated by 
  the green (blue) shaded areas in each case. The GW quasiparticle energies of the band extrema
  at high symmetry points are also shown, with blue squares and red circles indicating calculations 
  with or without semicore electrons, respectively. The connecting lines are guides to the eye.
  The coordinates of the high symmetry points in reciprocal lattice units are as follows: 
  $Z:(0,0,0.5)$, $X:(0.5,0,0)$, $Y:(0,0.5,0)$. 
  }
  \label{fig:bandstruct}
  \end {center}
  \end{figure*}

Figure~\ref{fig:bandstruct} shows the DFT/LDA band structures calculated using experimental
lattice parameters and without semicore electrons. Calculations including the semicore
states yield very similar band structures. For clarity we only show the dispersions
along the $Z$-$\Gamma$-$X$ path and along the $Y$-$\Gamma$ segment
running along the axis of the (A$_4$B$_6$)$_n$ ribbons.
The top of the valence band is found to be predominantly of S-3$p$ or Se-4$p$ character,
while the bottom of the conduction band comprises of Sb-5$p$ or Bi-6$p$ states,
consistently with previous calculations.\cite{Caracas2005, Nasr2011}

The band structures shown in Fig.~\ref{fig:bandstruct} exhibit several extrema
in proximity of the fundamental gap, making the direct and indirect transitions
almost degenerate. Table~\ref{tb:dftgap} shows that the energy separation between
direct and indirect DFT/LDA band gaps falls within 0.15~eV in all cases.
The data in the table suggest that in these compounds the direct transition will
most likely dominate over the indirect one, apart from a very narrow onset
of 0.1-0.2~eV. This observation is consistent with experimental evidence showing
a weak absorption onset just below the threshold for direct absorption.\cite{Zawawi1998,Versavel2007} 
Therefore for practical purposes, and in particular for photovoltaics applications,
stibnite, antimonselite, bismuthinite and guanajuatite can be considered 
as ``effectively direct gap'' semiconductors.

\subsection{Quasiparticle corrections}\label{sec.qp-corrections}

Figure~\ref{fig:bandstruct} shows that GW quasiparticle corrections lead to a moderate increase 
of the band gaps in all cases, while generally preserving the shape of the band extrema.
From this figure we deduce that a simple scissor operator should be able to capture
the most important effects of the GW corrections.

\begin{figure*}
  \begin{center}
  \includegraphics[width=0.8\textwidth]{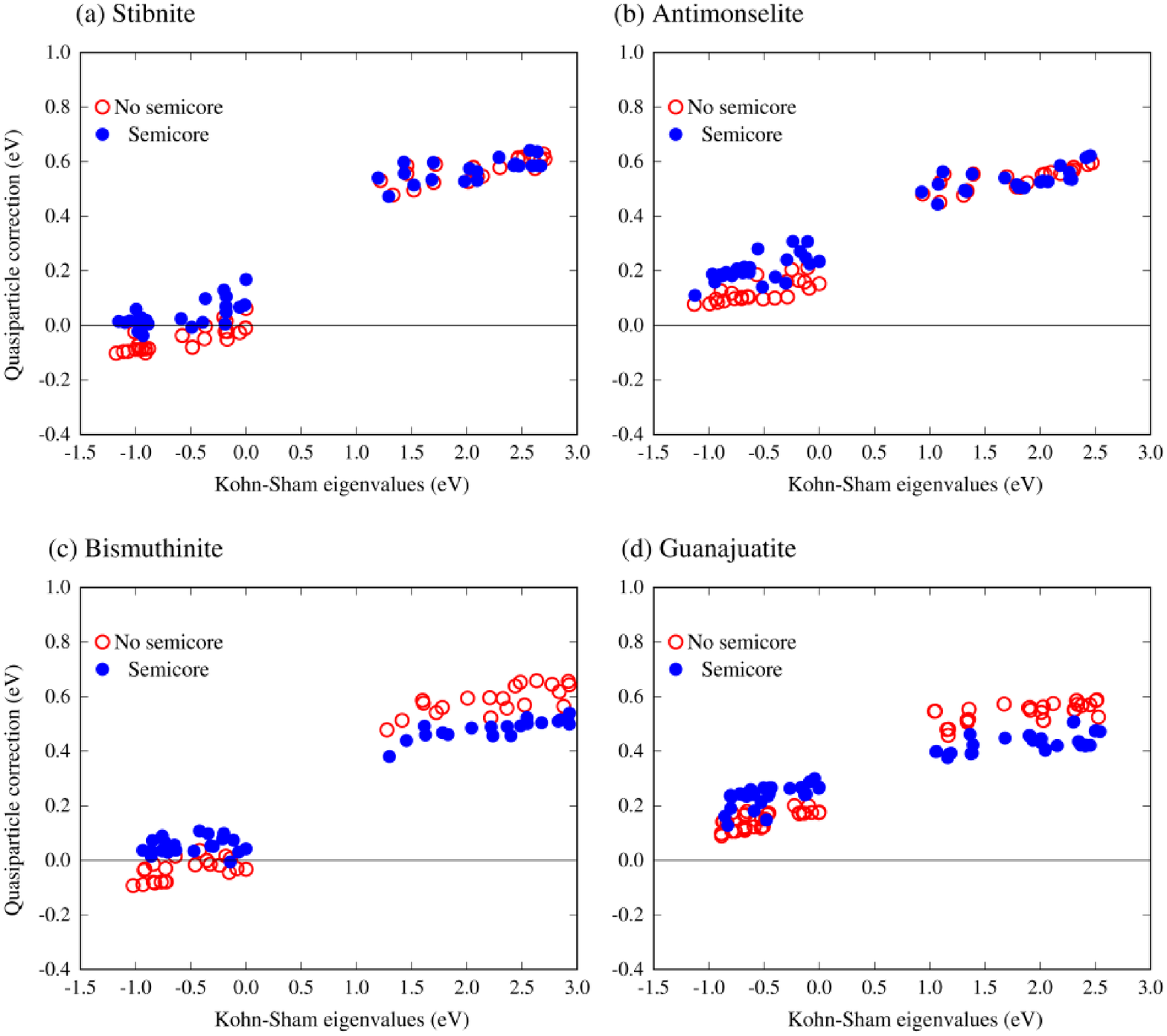}
  \caption{
  Quasiparticle corrections as a function of the corresponding DFT/LDA eigenvalues
  for (a) stibnite, (b) antimonselite, (c) bismuthinite, and (d) guanajuatite.
  Only eigenvalues at the high-symmetry points $\Gamma$, $X$ and $Z$ are considered.
  Blue disks and red circles indicate calculations with and without semicore electrons,
  respectively. All calculations were performed using experimental lattice parameters.
  }
  \label{fig:lev}
  \end{center}
  \end{figure*}

A detailed analysis of the quasiparticle corrections at the high-symmetry points $\Gamma$, $X$, 
and $Z$ is given in Fig.~\ref{fig:lev} and Table~\ref{tb:reslev}. In Fig.~\ref{fig:lev} we 
report the quasiparticle corrections as a function of the corresponding Kohn-Sham eigenvalues
around the band extrema.
In the cases of stibnite and antimonselite we observe that in the calculations with semicore electrons
the valence bands are slightly upshifted (by about 0.1~eV) as compared to calculations without semicore,
while the corrections to the conduction bands are essentially the same.
In the cases of bismuthinite and guanajuatite the effect of semicore is to shift the valence bands
up and the conduction bands down by a similar amount ($\sim$0.1~eV).
As a result of these small changes, the quasiparticle corrections to the band gaps 
calculated with or without semicore electrons can differ by up to 0.2~eV (cf.\ Table~\ref{tb:reslev}).

\begin{figure*}
  \begin{center}
  \includegraphics[width=0.8\textwidth]{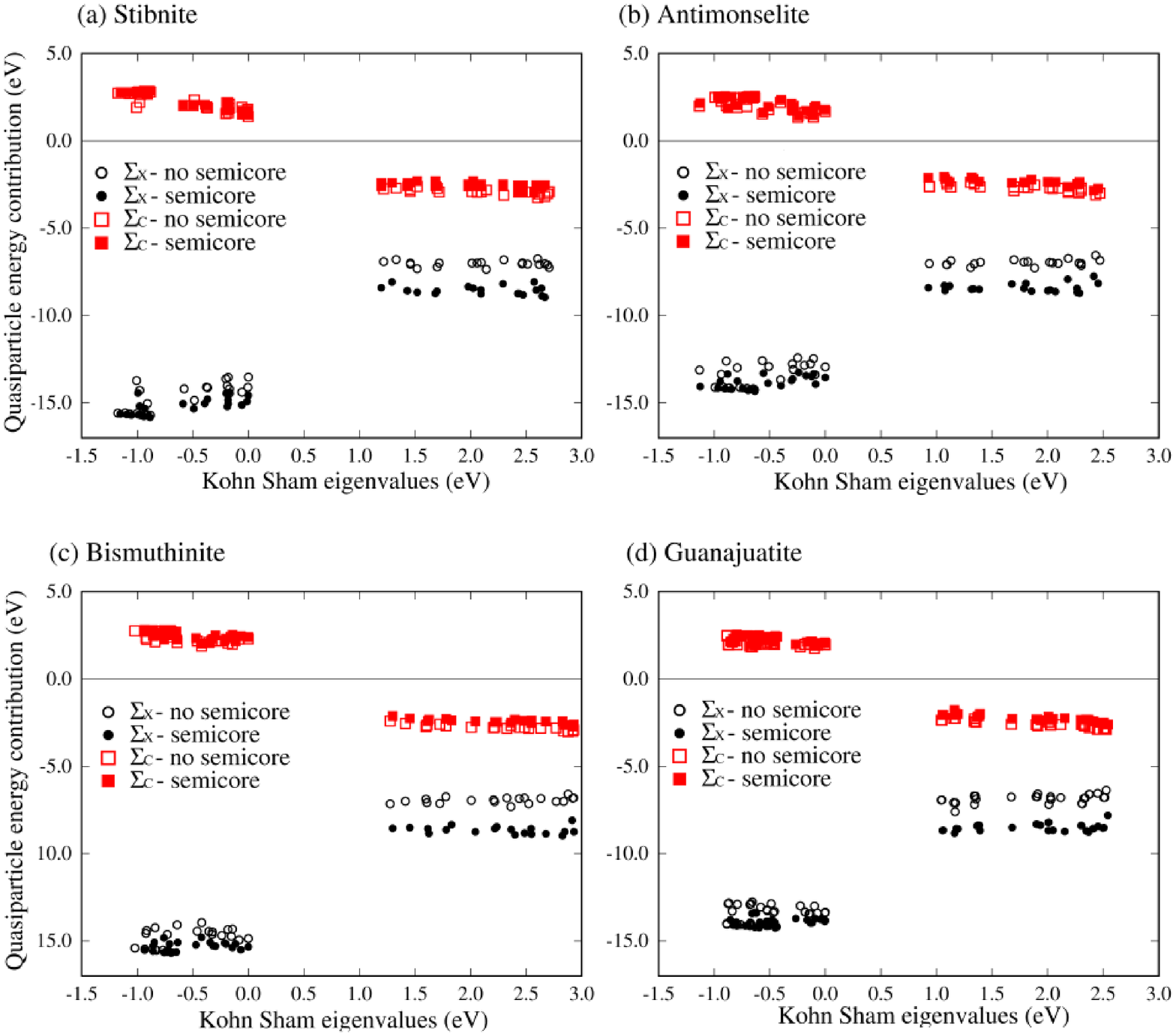}
  \caption{
  Exchange (black disks and circles) and correlation (red filled and empty squares) contributions
  to the quasiparticle corrections vs.\ DFT/LDA eigenvalues
  for (a) stibnite, (b) antimonselite, (c) bismuthinite, and (d) guanajuatite.
  Only eigenvalues at the high-symmetry points $\Gamma$, $X$ and $Z$ are considered.
  Filled and empty symbols indicate calculations with and without semicore states, respectively. 
  All calculations were performed using experimental lattice parameters.}
  \label{fig:sig}
  \end{center}
  \end{figure*}

Semicore electrons appear to slightly reduce the quasiparticle corrections as compared
to calculations without the semicore. This finding is consistent with previous 
calculations and can be rationalized as follows.\cite{Rohlfing1995, Tiago2004, Umari2012} 
The semicore $d$ states introduce additional contributions $\Sigma_x^{\rm SC}$ and $\Sigma_c^{\rm SC}$
to the GW self energy. Of these contributions, while the correlation part $\Sigma_c^{\rm SC}$
is small owing to the large energy separation between semicore states and conduction states,
the exchange part $\Sigma_x^{\rm SC}$ can be large since it does not contain energy 
denominators but is sensitive to the overlap between the band edge states and the semicore 
states. This interpretation is confirmed by Fig.~\ref{fig:sig}, where we can see
that the inclusion of semicore electrons does indeed affect the exchange part 
of the GW corrections, while at the same time the correlation component remains almost unchanged.

Table~\ref{tb:reslev} reports
the DFT/LDA eigenvalues and the corresponding quasiparticle corrections for the
valence band top and conduction band bottom at the high symmetry points $\Gamma$, $X$ and $Z$.
From this table we see that the LDA band gaps at these points are sensitive to the choice of the
lattice parameters, and this sensitivity is reflected in the corresponding quasiparticle
energies. Calculations performed using optimized lattice parameters or experimental
parameters can differ by up to 0.3~eV.
This observation may explain the lack of consensus between previous computational investigations
of the band structures of these compounds.\cite{Caracas2005, Patrick2011, Vadapoo2011rib, Koc2012}

Taken together the sensitivity of the quasiparticle energies to the presence
of semicore electrons and to the choice of lattice parameters leads to
non-negligible variations in the calculated band gaps.
This suggests that it is important to use some care 
when comparing the quasiparticle band structures of stibnite and related
compounds with experimental data.

In the remainder of this manuscript we will focus on calculations using
experimental lattice parameters and including semicore electrons, which
we consider our best estimates for the quasiparticle energies in these compounds.

\subsection{Relativistic corrections}
We calculate the relativistic corrections within DFT/LDA for all four structures using 
the experimental structure.
The corrections to the band edges at the high-symmetry points $\Gamma$, $X$ and $Z$
are reported in Table ~\ref{tb:spinorb}. 

In all four semiconductors the inclusion of spin-orbit coupling does not alter
the top of the valence band. This is consistent with the observation that
the states at the valence band top are predominantly associated with S or Se $p$ states.
On the other hand the bottom of the conduction bands are of Bi or Sb $p$ character
(see Fig.~\ref{fig:bandstruct}), hence a spin-orbit splitting is expected in this case. 
We calculate indeed a very small spin-orbit splitting for Sb$_2$S$_3$ and Sb$_2$Se$_3$,
which has the effect of lowering the conduction band minima by less than 0.1 eV. 
In the case of Bi$_2$S$_3$ and Bi$_2$Se$_3$ the spin-orbit splitting is as large
as 0.3-0.4~eV, consistent with the higher atomic number of Bi.

\subsection{Band gaps}

Table~\ref{tb:resgap} reports the quasiparticle band gaps calculated using the experimental
structures, including semicore electrons and relativistic corrections. 
The band gaps are obtained by considering the band extrema at $\Gamma$, $X$ and Z
and we give both the fundamental gap and the direct gap.
While in antimonselite, and bismuthinite the calculated minimum gap 
is indirect, the difference between direct and indirect gaps is within 0.1~eV.
In guanajuatite and stibnite the fundamental gap is direct. These results suggest that
all four compounds can be considered direct-gap semiconductors for practical
applications, especially in the area of optoelectronics.
The calculated direct gaps are 1.54~eV (stibnite), 1.27~eV (antimonselite),
1.42~eV (bismuthinite), and 0.91~eV (guanajuatite). As shown in Table~\ref{tb:resgap}
these values are in line with previous GW calculations where available,\cite{Vadapoo2011,Vadapoo2011rib}
and also rather close to measured optical gaps.

The comparison with experimental data is not straightforward since 
the experimental literature appears to only report optical gaps
(cf.\ literature review in Table~\ref{tb:resgap}).
However our calculations refer to quasiparticle gaps and do not include
excitonic effects. Including excitonic effects using the Bethe-Salpeter
approach\cite{Onida2002} would be rather challenging owing to
the large size of these systems.
To the best of our knowledge no excitonic effects were
measured or mentioned for any of the four compounds studied.
One exception is possibly the absorption spectrum reported in Ref.~\onlinecite{Versavel2007},
which exhibits some sharp features resembling excitonic peaks, however
the authors assigned those peaks to defects or internal reflections.
The agreement between our calculated quasiparticle gaps and the
measured optical gaps can be seen {\it a posteriori} as an indication
that excitonic shifts are small in this class of semiconductors.

\begin{figure}
  \begin{center}
  \includegraphics[width=0.9\columnwidth]{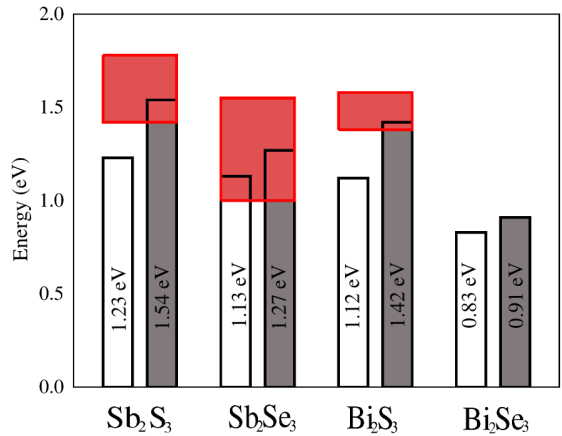}
  \caption{
  Schematic summary of the band gaps of stibnite, antimonselite, bismuthinite, and
  guanajuatite calculated in this work: Kohn-Sham gaps (empty rectangles) and GW gaps 
  (filled rectangles) including relativistic corrections.
   The band gaps were obtained by including semicore electrons 
  and using the experimental lattice parameters. The pink rectangles indicate the
  range of experimental optical gaps reported in Table~\ref{tb:resgap}}.
  \label{fig:stat}
  \end{center}
  \end{figure}

Figure~\ref{fig:stat} provides a schematic view of our final calculated band gaps
(GW + SOC) compared to the Kohn Sham band gaps (DFT/LDA + SOC) and experiment.
  \begin {table}
  \begin {center}
  \begin {tabular}{lrrrrrrrr}
  \hline
  \hline
  \\[-8pt]
  & \multicolumn{4}{r}{Optimized parameters} & \multicolumn{4}{r}{Expt. parameters} \\
  \\[-8pt]
  & \multicolumn{2}{c}{LDA} & \multicolumn{2}{c}{GW} & \multicolumn{2}{c}{LDA} & \multicolumn{2}{c}{GW} \\
  & w/o S \phantom{a}& S
  & w/o S \phantom{a}& S 
  & w/o S \phantom{a}& S 
  & w/o S \phantom{a}& S \\
  \\[-8pt]
  \hline
  \\[-8pt]
  &\multicolumn{8}{c}{Stibnite}\\
  \\[-8pt]
  $\Gamma_v$& 0.00  & 0.00 &  0.10  & 0.20 & 0.00  & 0.00 & 0.06 &  0.17 \\
  $\Gamma_c$& 1.15  & 1.11 & 1.58  & 1.52  & 1.33  & 1.29 & 1.81 &  1.77 \\
  $X_v$ & -0.05 &  -0.03 & -0.03 &  0.09 & 0.00  & -0.01 & -0.02 &  0.06 \\
  $X_c$ & 1.40 &  1.39 & 1.90 &  1.88 & 1.46 &  1.43 & 2.01 &  1.99 \\
  $Z_v$ & -0.16 &  -0.14 & -0.14  & -0.04 & -0.06  & -0.06 &  -0.08  & 0.01 \\
  $Z_c$ & 1.17 &  1.17 & 1.65 &  1.65 & 1.22 &  1.20 &  1.75 &  1.74 \\
  \\[-8pt]
  &\multicolumn{8}{c}{Antimonselite}\\
  \\[-8pt]
  $\Gamma_v$& -0.12  & -0.12 &  0.07  & 0.18  & -0.11 & -0.11  & 0.11  & 0.20 \\
  $\Gamma_c$& 0.97  & 0.91 &  1.40  & 1.32 &  1.09 &  1.07 &  1.54  & 1.52 \\ 
  $X_v$     & 0.00  & 0.00 &  0.19  & 0.29 &  0.00 &  0.00 &  0.15  & 0.23 \\
  $X_c$     & 1.05  & 1.00 &  1.53  & 1.46 &  1.10 &  1.08 &  1.62  & 1.60 \\
  $Z_v$     & -0.23  & -0.24  & -0.06  & 0.02 & -0.09  & -0.09 & 0.04  & 0.14 \\
  $Z_c$     & 0.92  & 0.91 &  1.37 & 1.35 &  0.94 &  0.93 &  1.42  & 1.42 \\
  \\[-8pt]
  & \multicolumn{8}{c}{Bismuthinite}\\
  \\[-8pt]
  $\Gamma_v$& -0.10 &  -0.04 &  -0.04  & 0.08  & -0.14  & -0.12 &  -0.14  & -0.04 \\
  $\Gamma_c$& 1.14 &  1.14 &  1.57 &  1.48 &  1.28 &  1.30 &  1.76 & 1.68 \\
  $X_v$     & 0.00 &  0.00 &  0.02 &  0.09 &  0.00 &  0.00 &  -0.03 &  0.04 \\
  $X_c$     & 1.50 &  1.67 &  2.04 &  2.09 &  1.61 &  1.63 &  2.18 &  2.09 \\
  $Z_v$	  & -0.15 &  -0.11& -0.13 &  -0.04  & -0.08 &  -0.07 &  -0.11  & -0.04 \\
  $Z_c$   & 1.43  & 1.51 &  1.89  & 1.93 &  1.41 &  1.45 &  1.93  & 1.89 \\
  \\[-8pt]
  & \multicolumn{8}{c}{Guanajuatite}\\
  \\[-8pt]
  $\Gamma_v$& -0.02 &  -0.02 &  0.18  & 0.30  & -0.07  & -0.04  & 0.11  & 0.26 \\
  $\Gamma_c$& 0.95 &  0.89 &  1.39  & 1.24 &  1.17 &  1.16 &  1.63 &  1.54 \\
  $X_v$     & 0.00 &  0.00 &  0.19  & 0.28 &  0.00 &  0.00 &  0.18 &  0.27 \\
  $X_c$     & 1.07 &  1.18 &  1.61  & 1.58 &  1.04 &  1.06 &  1.59 &  1.45 \\
  $Z_v$	  & -0.24 &  -0.19&  -0.06  & 0.05  & -0.14  & -0.12  & 0.04  & 0.12 \\
  $Z_c$	  & 1.17 &  1.25 &  1.63 &  1.65 &  1.16 &  1.18 &  1.64 &  1.57 \\
  \\[-8pt]
  \hline
  \hline
  \end{tabular}
  \caption{
  Quasiparticle energies of stibnite, antimonselite, bismuthinite and guanajuatite
  at the high-symmetry points $\Gamma$, $X$, $Z$ vs.\ the corresponding
  DFT/LDA eigenvalues. We report both sets of results obtained using optimized
  or experimental lattice parameters. The columns labelled ``S'' and
  ``w/o S'' indicate calculations with and without semicore electrons, respectively.
  For each high-symmetry point we consider the energies at the valence band top 
  (e.g.\ $\Gamma_v$) and the conduction band bottom (e.g.\ $\Gamma_c$).
  All values are in units of eV.
  }
  \label{tb:reslev}
  \end {center}
  \end {table}

  \begin {table*}
  \begin {center}
  \begin {tabular}{l rrrrrrrrrrrrrrrr}
  \hline
  \hline
  & \multicolumn{4}{c}{Sb$_2$S$_3$} 
  & \multicolumn{4}{c}{Sb$_2$Se$_3$}
  & \multicolumn{4}{c}{Bi$_2$S$_3$}
  & \multicolumn{4}{c}{Bi$_2$Se$_3$} \\
  \\[-8pt]
   & w/o SOC && SOC &
   & w/o SOC && SOC & 
   & w/o SOC && SOC &
   & w/o SOC && SOC &\\ 
  \hline
  \\[-8pt]

  $\Gamma_v$& 0.00 &&  0.00 &
  			& -0.11 && 0.00 &
  			& -0.12 && -0.02&
  			& -0.04 && -0.01&\\
  $\Gamma_c$& 1.29 && -0.06 &
  			& 1.07 && -0.05 &
  			& 1.30 && -0.32 &
  			& 1.16 && -0.38 &\\
  $X_v$ & -0.01 && 0.00 &
  		& 0.00 && 0.00  &
  		& 0.00 && -0.02 & 
  		& 0.00 && -0.02 &\\
  $X_c$ & 1.43 &&  -0.04 &
  		& 1.08 && -0.03  &
  		& 1.63 && -0.40  &
  		& 1.06 && -0.27  &\\
  $Z_v$ & -0.06 && 0.00 &
  		& -0.09 && 0.00  &
  		& -0.07 && -0.02&
  		& -0.12 && -0.01&\\
  $Z_c$ &  1.20 && -0.02&
  		& 0.93 && -0.02 &
  		& 1.45 && -0.31 & 
  		& 1.18 && -0.28 &\\
  \hline
  \hline
  \end{tabular}
  \caption{
  Relativistic corrections calculated for stibnite, antimonselite,  
  bismuthinite and guanajuatite at high-symmetry points. The column labelled ``w/o SOC''
  indicates the scalar relativistic values of the band edges, while the column labelled ``SOC''
  reports the corresponding relativistic corrections.
  All calculations are performed using the experimental structures and including
  semicore $d$ states. All values are in units of eV.}
  \label{tb:spinorb}
  \end {center}
  \end {table*}

  \begin{table*}
  \begin{center}
  \begin{tabular}{ll l l l l l l l l}
\hline 
\hline
   & &\multicolumn{1}{c}{Previous} & \multicolumn{1}{c}{Present} 
   && \multicolumn{1}{c}{Previous} & &\multicolumn{1}{c}{Present}  & & \multicolumn{1}{c}{Experiment} \\
   & &\multicolumn{1}{c}{DFT} & \multicolumn{1}{c}{DFT+SOC} 
   && \multicolumn{1}{c}{GW} & &\multicolumn{1}{c}{GW+SOC}  & & \\
\hline
  Sb$_2$S$_3$ &&  1.55$^a$, 1.76$^b$, 1.3$^c$, 1.18$^d$, 1.22$^e$ & 1.23 && 1.67$^e$ && 1.54 & &1.73$^f$, 1.42-1.65$^g$, 1.78$^h$, 1.7$^i$, 1.74$^j$ \\
  Sb$_2$Se$_3$ && 1.14$^a$, 0.99$^d$, 0.79$^k$, 0.89$^e$ & 1.13 (0.91) && 1.21$^k$ && 1.27 (1.17) && 1.55$^l$, 1.2$^i$, 1.0 - 1.2$^m$ \\
  Bi$_2$S$_3$ && 1.47$^a$, 1.32$^n$, 1.63$^n$, 1.45$^n$, 1.67$^n$ & 1.12 (1.00) &&  && 1.42 (1.34) && 1.4$^o$, 1.38$^p$, 1.58$^{q,j}$\\
  Bi$_2$Se$_3$ && 0.9$^a$, 1.1$^r$ & 0.83 &&  && 0.91 && \\
\hline
\hline
  \multicolumn{6}{l}{\footnotesize{$^a$ Ref.~\citenum{Caracas2005}, $^b$ Ref.~\citenum{Nasr2011}, $^c$ Ref.~\citenum{Patrick2011}, $^d$ Ref.~\citenum{Koc2012}, $^e$ Ref.~\citenum{Vadapoo2011rib}, $^f$ Ref.~\citenum{Versavel2007},}}\\
  \multicolumn{6}{l}{\footnotesize{$^g$ Ref.~\citenum{Mahanty1997}, $^h$ Ref.~\citenum{Zawawi1998}, $^i$ Ref.~\citenum{Black1957}, $^j$ Ref.~\citenum{Yesugade1995},$^k$ Ref.~\citenum{Vadapoo2011}, $^l$ Ref.~\citenum{Torane1999}, $^m$ Ref.~\citenum{Rodriguez-Lozcano2005},}}\\
  \multicolumn{6}{l}{\footnotesize{$^n$ Ref.~\citenum{Sharma2012}, $^o$ Ref. ~\citenum{Gildart1961}, $^p$ Ref.~\citenum{Pradeep1991}, $^q$ Ref.~\citenum{Mahmoud1997}, $^r$ Ref.~\citenum{Sharma2010}}}\\
\end{tabular}
  \caption{Comparison between calculated and measured band gaps of stibnite, antimonselite, bismuthinite, 
  and guanajuatite. We report the direct band gaps calculated within DFT/LDA and GW
  after the SOC corrections, as well as the measured optical gaps. The direct gaps are reported for the
  $\Gamma$ point. The values in parentheses indicate the calculated indirect band gaps in each case. 
  All values are in units of eV.
  Our calculations include semicore electrons and are performed using the experimental structures.
  }
  \label{tb:resgap}
  \end{center}
  \end{table*}

\section{Discussion}\label{sec:Discussion}

Taking the calculated quasiparticle band gaps of 0.9-1.5~eV   
as representative of the optical gaps,
the four semiconductors considered here lie precisely in the range of the optimal Shockley-Queisser 
performance.\cite{Shockley1960} 
The Shockley-Queisser analysis addresses the ultimate efficiency of a solar cell based on a single
material as light absorber and electron conductor, e.g.\ silicon solar cells. In this analysis
the optimum efficiency results from a trade-off between maximizing the band gap
in order to increase the photovoltage, and minimizing the band gap in order to increase
the photocurrent.\cite{Shockley1960}

\begin{figure}
  \begin{center}
  \includegraphics[width=0.9\columnwidth]{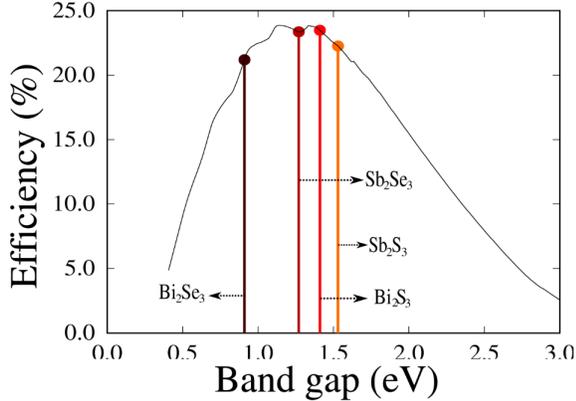}
  \caption{
  Ideal efficiency of nanostructured solar cells based on semiconductors of the
  stibnite family. The theoretical efficiency as a function of the band gap energy
  (black solid curve) is calculated using the prescription of Ref.~\onlinecite{Snaith2010} with a
  loss-in-potential of 0.3~eV and a fill factor of 73\%.
  }
  \label{fig:efficiency}
  \end{center}
  \end{figure}

In the case of nanostructured solar cells based on the donor/acceptor concept such as for
instance semiconductor-sensitized solar cells,\cite{Hodes2008,Chang2010} the Shockley-Queisser
analysis needs to be modified in order to take into account the energy-level alignment
at the donor/acceptor interface. In fact, while the photocurrent is still determined
by the optical gap of the absorber (typically the donor), at variance with conventional
bulk solar cells the photovoltage is dictated by the difference between the lowest
unoccupied states of the acceptor and the highest occupied states of the donor.
This effect can be taken into account by introducing the concept of ``loss-in-potential'',\cite{Snaith2010}
which is the reduction of the photovoltage resulting from the energy mismatch and additional
losses. Loss-in-potentials estimated for actual devices can be as large as $\sim 1$~eV,
and the most optimistic scenario would correspond to losses as small as 0.3~eV.\cite{Snaith2010}
Figure~\ref{fig:efficiency} shows the theoretical efficiency of semiconductor-sensitized
solar cells based on stibnite, antimonselite, bismuthinite, and guanajuatite,
calculated using the prescription of Ref.~\onlinecite{Snaith2010} for a loss-in-potential of 0.3~eV.
While these estimates are very crude and the projections are possibly too optimistic, 
it is interesting to note that all of these four materials cluster very near the optimum 
power conversion efficiency of 20-25\%.

From Fig.~\ref{fig:efficiency} we infer that the four compounds studied here are all promising
candidate for nanostructured photovoltaic applications, with antimonselite and bismuthinite 
slightly superior to stibnite. In particular it cannot be excluded that guanajuatite, 
even if unstable at room temperature in bulk form, could be stabilized as a nanostructure. 
Given its projected maximum efficiency in Fig.~\ref{fig:efficiency}, it might be worth 
to attempt the synthesis of guanajuatite nanoparticles.
In the case of bismuthinite Refs.~\onlinecite{Patrick2011,Lutz2012} showed that
this material does not work as a semiconductor sensitizer for TiO$_2$, owing to the
incorrect energy-level alignment at the interface. However it cannot be excluded
that bismuthinite could still reach the ideal efficiency when combined with an
alternative acceptor, e.g.\ SnO$_2$ or ZnO.

\section{Conclusions}\label{sec:Conclusions}

In this work we report a systematic study of the quasiparticle band structures of the four isostructural 
metal chalcogenides stibnite (Sb$_2$S$_3$), antimonselite (Sb$_2$Se$_3$), bismuthinite
(Bi$_2$S$_3$), and guanajuatite (Bi$_2$Se$_3$), within the GW approximation.

In order to ensure reproducibility of our results we have placed an emphasis on convergence
tests and explored the effects of various calculation parameters, such as for instance the
role of semicore $d$ electrons and lattice parameters.
The inclusion of semicore electrons in the calculations is found to modify
the band gaps by up to 0.2~eV, and the choice of experimental vs.\ optimized
lattice parameters can lead to differences of up to 0.3~eV in the calculated gaps.
These findings indicate that some caution should be used in discussing the theoretical
band gaps of these materials and in comparing with experiment.
Relativistic effects are found to lower the conduction bands of all four materials.
Spin-orbit coupling effects are important in Bi$_2$S$_3$ and Bi$_2$Se$_3$, where they
reduce the band gaps by 0.3-0.4 eV, while they are essentially negligible for Sb$_2$S$_3$ and Sb$_2$Se$_3$.

Our calculations indicate that all four compounds have direct band gaps, barring indirect transitions marginally
below the direct gap. The calculated band gaps are 1.54~eV (stibnite),
1.27~eV (antimonselite), 1.42~eV (bismuthinite) and 0.91~eV  (guanajuatite).
These values fall within the range of measured optical gaps, although it must be observed that
there is a considerable scatter in the experimental data, possibly due to different
preparation conditions.

Using a modified Shockley-Queisser analysis,\cite{Snaith2010}
we estimate the ultimate performance of solar cells based on these compounds as
light sensitizers.
This analysis indicates that all four materials have potential for high-efficiency
nanostructured solar cells. The highest theoretical efficiencies are obtained for
antimonselite and bismuthinite, followed closely by stibnite and guanajuatite, the high temperature
polymorph of the topological insulator Bi$_2$Se$_3$.

Future calculations should address the optical absorption spectra of these compounds
within the Bethe-Salpeter approach, in order to establish whether excitonic effects
are small as our data appear to suggest. It will be also interesting to extend
the present study to the case of individual nanoribbons of these metal chalcogenides,
since liquid-phase exfoliation techniques for van der Waals bonded materials are becoming
increasingly popular.\cite{Nicolosi2011}

We hope that the present study will contribute to the ongoing research
on new materials for energy applications, and stimulate further efforts to 
understand and exploit these fascinating and relatively unexplored compounds.

\begin{acknowledgements}
This work is supported by the UK EPSRC and the ERC under EU FP7/ERC Grant No. 239578.  Calculations 
were performed at the Oxford Supercomputing Centre. Figures rendered using \texttt{Xcrysden}.\cite{Kokalj2003}
\end{acknowledgements}

\end {document}